\documentclass[12pt,a4paper]{article}
\usepackage{amsmath}
\usepackage{amsfonts}
\usepackage{amssymb}
\usepackage{graphicx}
\usepackage{float}
\newcommand{\bec}{\begin{center}}
\newcommand{\ec}{\end{center}}
\usepackage{feynmp-auto}
\usepackage[left=2cm,right=2cm,top=2cm,bottom=2cm]{geometry}
\title{Computer modeling of production cross sections for beyond SM particles}
\author{Tetiana Obikhod, Ievgenii Petrenko}
\date{%
    \it{Kyiv Institute for Nuclear Research NAS of Ukraine}\\%
    \today
}
\begin{document}

\maketitle 

\section{Abstract}
\par
In the paper is presented computer modelling of BSM physics, such as Dark matter candidates, di-jet resonances and new Higgs bosons with the help of MCFM program.

\section{Introduction}
\par
The search for new physics is one of the main goals of the LHC experiment. Since the most of the experimental data obtained at the LHC are well described by SM, the search for physics beyond SM (BSM) becomes more precision. In this aspect, the studied physical processes and the corresponding computer software are of particular importance. 
\par
Numerous experimental searches for di-jet resonances, such as string resonances, scalar diquarks, axigluons and colorons, excited quarks, color-octet scalars, Kaluza-Klein partners of W and Z bosons, Randall-Sundrum (RS) Gravitons, and dark matter (DM) mediators are predicted by a variety of new physical models, \cite{CMS-PAS-EXO-19-012}. As for the candidate of DM, this question is of particular interest, because the study of the nature of DM is one of the most important issue of today. Despite the fact that the existence of DM has now been well established \cite{Jarosik_2011}, its nature is still unknown. There are many candidates of DM, in particular, weakly-interacting massive particles (WIMPs) which are classified as "cold", "warm", or "hot". One of the motivated DM model is a new neutral particle with mass value of the order of the weak scale, \cite{PhysRevD.87.054030}, as a mediator which couples to SM particle with the production of a pair of fermionic DM particles.
\par
Another searches for BSM physics are connected with physics of Higgs boson. The observation of a Higgs boson with a mass of 125 GeV by the ATLAS and CMS experiments \cite{20121, 201230} confirms SM predictions. The anomalous interaction of the Higgs boson with the top quark, has been experimentally studied through the measurement of the Higgs boson production in association with a top quark, \cite{ATLAS-CONF-2019-004}. Recent ATLAS Higgs results using Run-2 data at a center-of-mass energy of 13 TeV with up to an integrated luminosity of 139 fb$^{-1}$ is observed in the diphoton decay mode with a significance of 4.9 standard deviations relative to the background-only hypothesis. To probe BSM physics, coupling of Higgs boson with top quark showed that Higgs boson will continue to provide an important probe for new physics. Furthermore, measuring of the interaction of the Higgs boson and the top-quark, sheds light on the instability of the electroweak vacuum, \cite{ALEKHIN2012214}. Extensions of the SM, such as the Minimal Supersymmetric Standard Model (MSSM) \cite{WESS197439, DIMOPOULOS1981150} and two-Higgs-doublet model (2HDM) \cite{BRANCO20121} predict new spin-0 states, such as additional scalar (H) or pseudoscalar (A) Higgs bosons. Therefore, the study of the properties of these particles may be associated with new BSM physics.
\par
In our paper we have considered the following particles and processes:\\
$\bullet$ Di-jet resonances;\\
$\bullet$ Studdying the properties of CP-odd A boson;\\
$\bullet$ Higgs boson production process in association with pair of top quarks;\\
$\bullet$ Dark matter production processes.

For this purpose we used program MCFM - Monte Carlo for FeMtobarn processes, v.9.0, \cite{campbell2019precision}. MCFM is a parton-level program that gives LO (leading order) and NLO (next-to-leading-order) predictions for the wide range of processes at the LHC. The specific processes presented above are implemented at LO in the MCFM program. We have used the latest, version 9.0 of MCFM program, which accounts PDF uncertainties and effectively takes into account QCD scale.

\section{Calculations for production cross sections}
\subsection{Di-jet resonances}
Hadronic collisions of partons in $2\rightarrow2$ scattering processes are described by QCD at small particle ejection angles with respect to the direction of the initial partons. These hadronic collisions are accompanied by the formation of a short-lived di-jet resonant characterized by smooth and monotonically decreasing distribution for the di-jet invariant mass, M$_ij$. BSM theories predict the existence of di-jet resonance states, which decay into two jets at large polar angles. Di-jet resonances has been searched for during long period of time, \cite{1988127, bauce2017search}. Recent experimental data, \cite{CMS-PAS-EXO-19-004, CMS-PAS-EXO-19-012} connected with searches for such exotic objects as DM candidates, RS Gravitons, Kaluza-Klein partners of gauge bosons W, Z  served as an occasion for further investigation and computer modelling of di-jet properties at higher energies. 
\par
\begin{figure}[htbp]
\bec
 \includegraphics[width=0.71\textwidth]{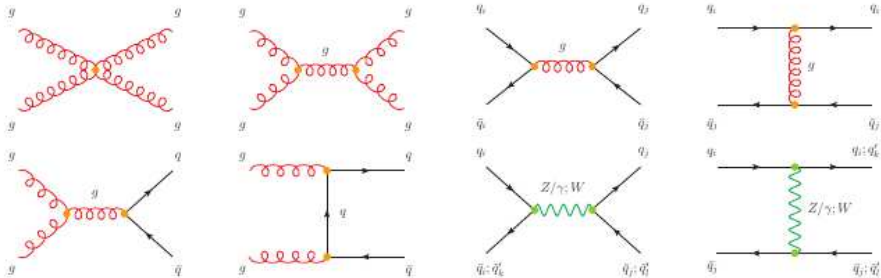}
 %scale=0.85
 \caption{\label{fig:sample_proc} Sample of di-jet processes, from \cite{PhysRevD.94.093009}}
\ec
\end{figure}

Di-jet productions via QCD and EW interactions are shown in fig.\ref{fig:sample_proc}.

\begin{figure}[htbp]
\bec
 \includegraphics[width=0.41\textwidth]{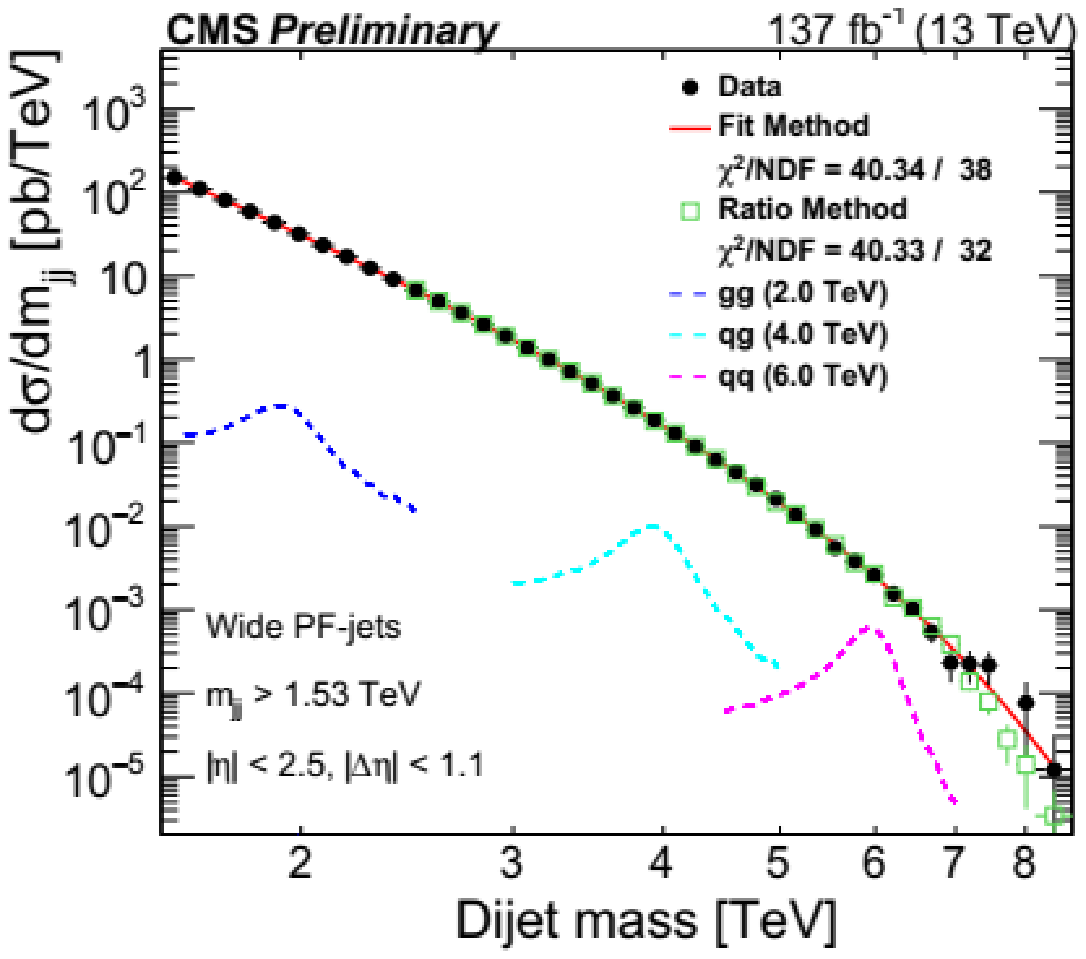}
 \includegraphics[width=0.48\textwidth]{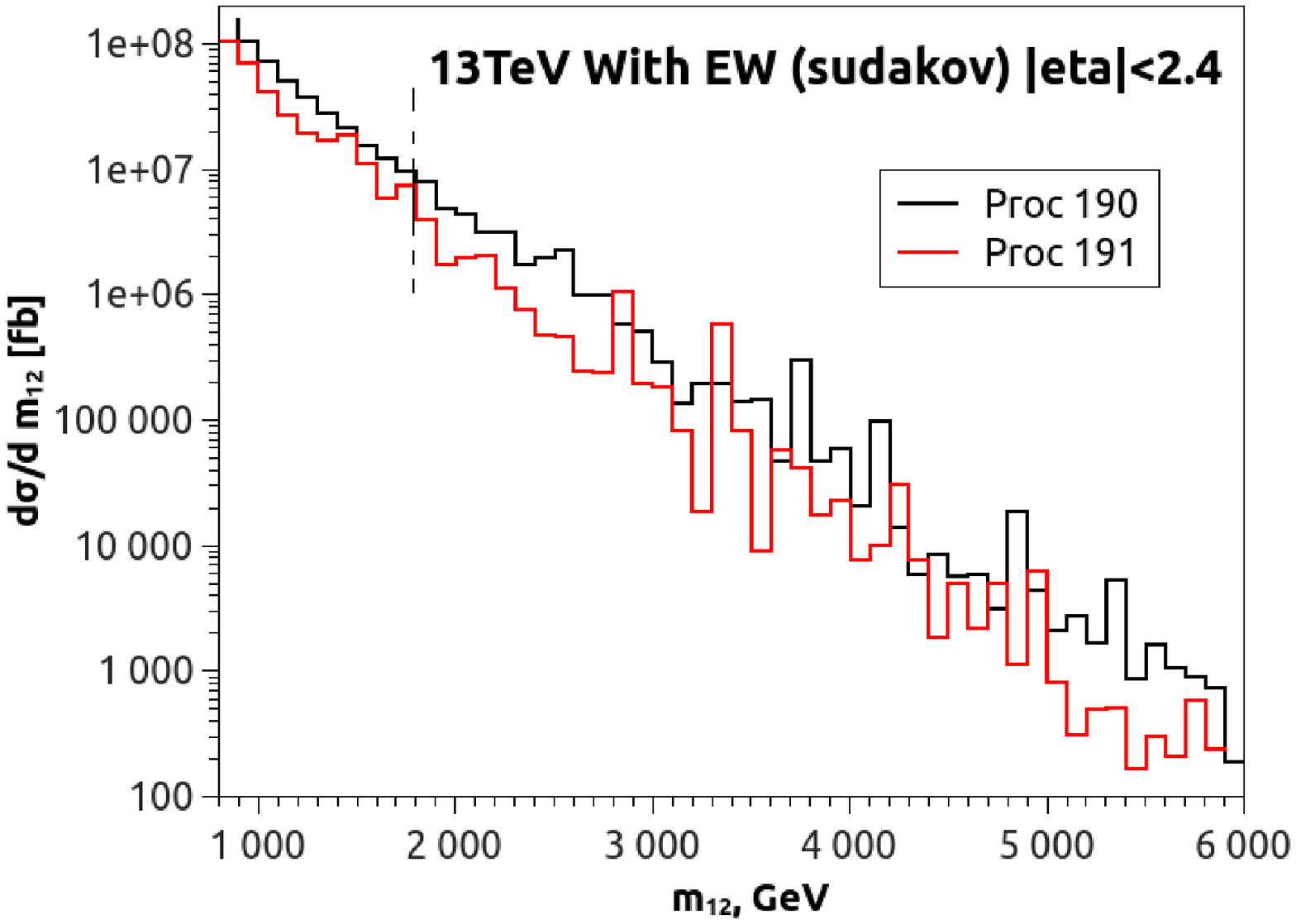}
 \caption{\label{fig:dijet_comparison} Experimental di-jet mass spectrum (left) compared to modeled di-jet spectrum (right) for two processes: 190 – only strong interactions and 191 interactions with EW one-loop corrections at 13 TeV.}
\ec
\end{figure}

We'll consider the data of experiment \cite{CMS-PAS-EXO-19-012} with mass of resonances greater than 1.8 TeV decaying to a pair of jets, obtained in proton-proton collisions at $\sqrt{s}$ = 13 TeV with integrated luminosity of 137 $fb^{-1}$. The comparision of data with our calculations at the same energy of 13 TeV is presented in fig.\ref{fig:dijet_comparison}.

We added electroweak (EW) corrections (191 process with Sudakov label) as at high energies they play a significant role due to the occurrence of soft and collinear radiation of virtual and real W and Z bosons, which gives rise to Sudakov-like corrections, \cite{PhysRevD.94.093009}. Comparision of data with inclusion of the experimental di-jet mass limitations ($>$ 1.8 TeV) shows the difference of less or about 1 order with left part of fig.\ref{fig:dijet_comparison}. 

\par
\begin{figure}[htbp]
\bec
 \includegraphics[width=0.55\textwidth]{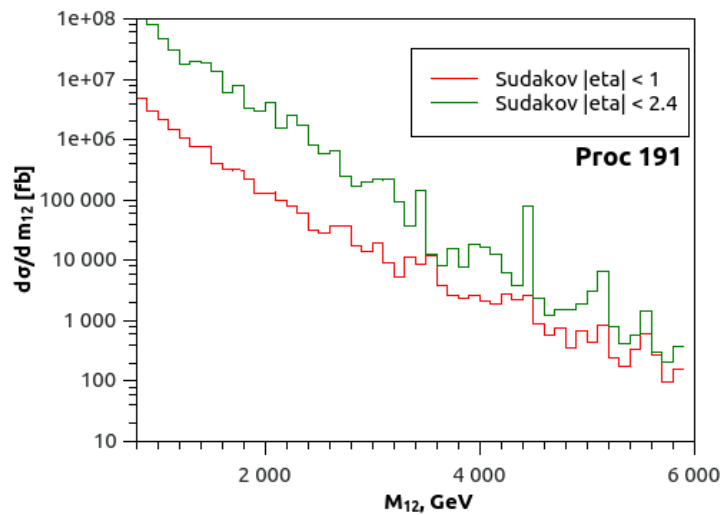}
 \caption{\label{fig:dijet_calc_14} Di-jet mass spectrum for different pseudorapidity regions and with EW corrections at 14 TeV.}
\ec
\end{figure}

Investigation of the effect on kinematic distributions such as the invariant mass of the lepton pair leads to the necessity of the inclusion of weak one-loop corrections to the total rate for the Neutral-Current DY process. The effect of the weak one-loop corrections on the lepton rapidity distributions is rather mild, since they are not very sensitive to the presence of the weak Sudakov logarithms. As the exact EW and Sudakov corrections are in good agreement inside the very central rapidity region, \cite{PhysRevD.94.093009}, we calculated di-jet mass spectrum for different pseudorapidity regions, presented in fig.\ref{fig:dijet_calc_14}.

From fig.\ref{fig:dijet_calc_14} we see the increase of the production cross section with larger pseudorapidity, but in the region of $M_{ij}$ from 4 to 6 TeV we see the smooth growth of the second one. 
\par
Angular distributions of di-jet processes are also of great interest for futher experimental searches. We calculated $d\sigma/d\eta$ for two processes, 190 and 191 at 14 TeV.
\newpage
\begin{figure}[htbp]
\bec 
 \includegraphics[width=0.45\textwidth]{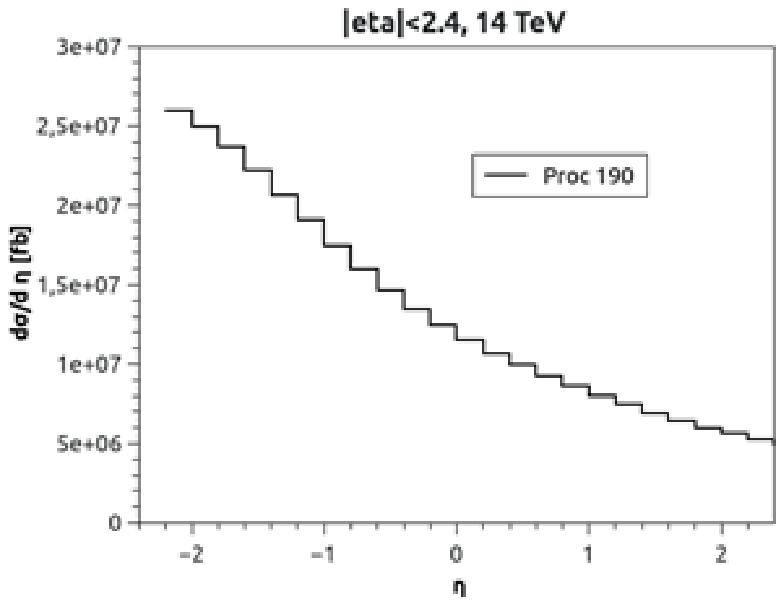}
 \includegraphics[width=0.45\textwidth]{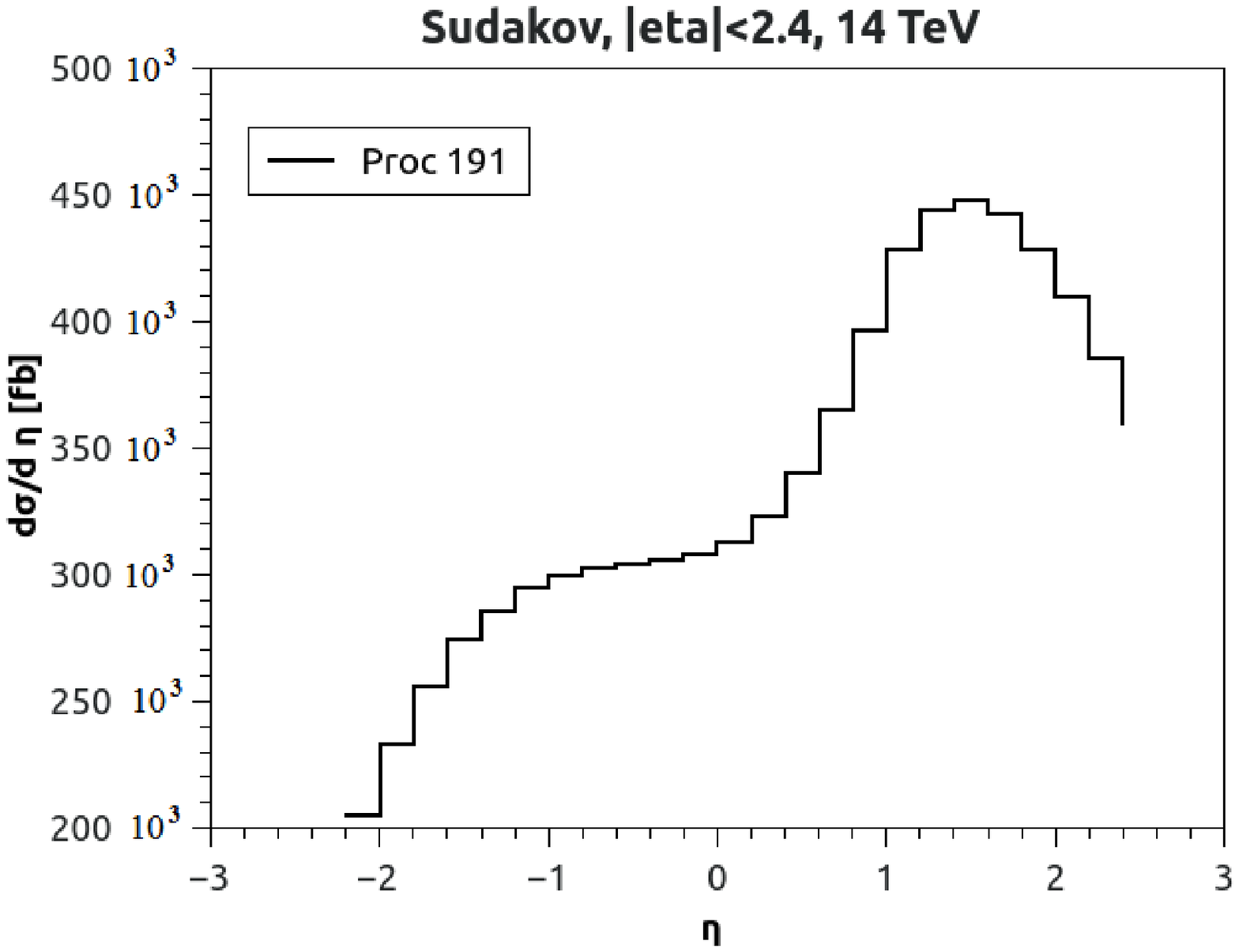}
 \caption{\label{fig:dijet_calc_eta} Differential production cross section as function of pseudorapidity calculated for two processes: left – 190 (only strong interactions), right – 191 (with inclusion of Sudakov EW corrections) at 14 TeV.}
\ec
\end{figure}

\begin{figure}[htbp]
\bec
 \includegraphics[width=0.45\textwidth]{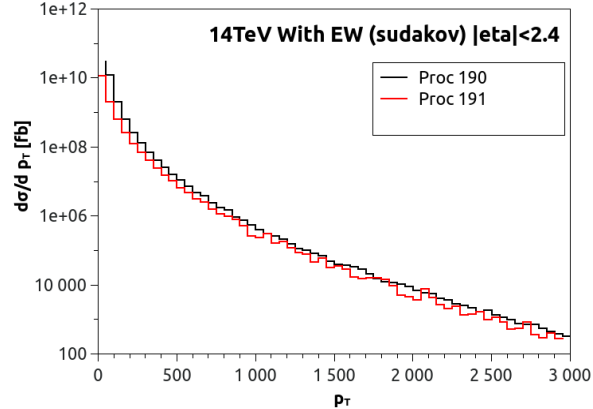}
 \caption{\label{fig:dijet_calc_pt} Differential production cross sections with respect to the di-jet transverse momentum $p_T$.}
\ec
\end{figure}

	From fig.\ref{fig:dijet_calc_eta} we see the large difference in angular distributions of these two processes. The character of the function in the left part signals about the predominance of di-jet direction backward to the axis of proton-proton interactions. The right side of the figure represents a sharp increase in function at about 14$^0$ and 20$^0$ to the direction of proton-proton collision. In fig.\ref{fig:dijet_calc_pt} another important characteristics is presented -- contribution of differential production cross section with respect to the transverse momentum of two jets.\\

From fig.\ref{fig:dijet_calc_pt} we see the smooth behavior of a curve that decreases with increasing $p_T$ but sharply increasing at small transverse momenta.

\subsection{Pseudoscalar boson production}
The detailed study of the production and decay modes of the new particle with mass of 125 GeV at the LHC indicates that the new particle is indeed compatible with the SM Higgs boson. Nevertheless, many scenarios of physics beyond SM include a SM-like Higgs boson as part of an extended sector of scalar particles. In any case, searches for new Higgs bosons are connected with the measurements of the properties of new particles of an extended Higgs boson sector. One of such particle is pseudoscalar Higgs boson, A, predicted by MSSM model. The Higgs boson couples to a pair of gluons via a loop of heavy fermions (top or bottom quarks) 
\begin{center}
\includegraphics[width=0.35\textwidth]{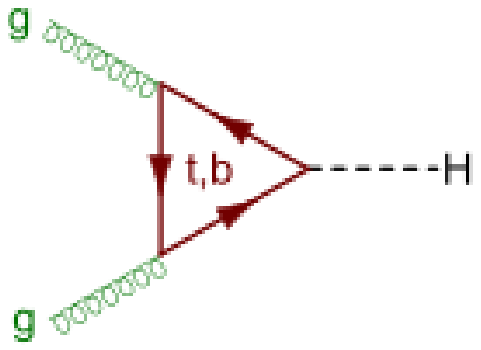}
\end{center}
which is accounted by the inclusion of the matrix elements with  dependence on the top quark mass. The calculation can only be performed at LO. We considerd two processes of MCFM v.9 program:

\[206  \ \ f(p_1)+f(p_2) \rightarrow A (b(p_3)+\overline{b}(p_4)) + f(p_5) ;\]
\[207 \ \ f(p_1)+f(p_2) \rightarrow A (\tau^{-}(p_3)+\tau^{+}(p_4)) + f(p_5) .\]

\begin{figure}[htbp]
\bec
 \includegraphics[width=0.45\textwidth]{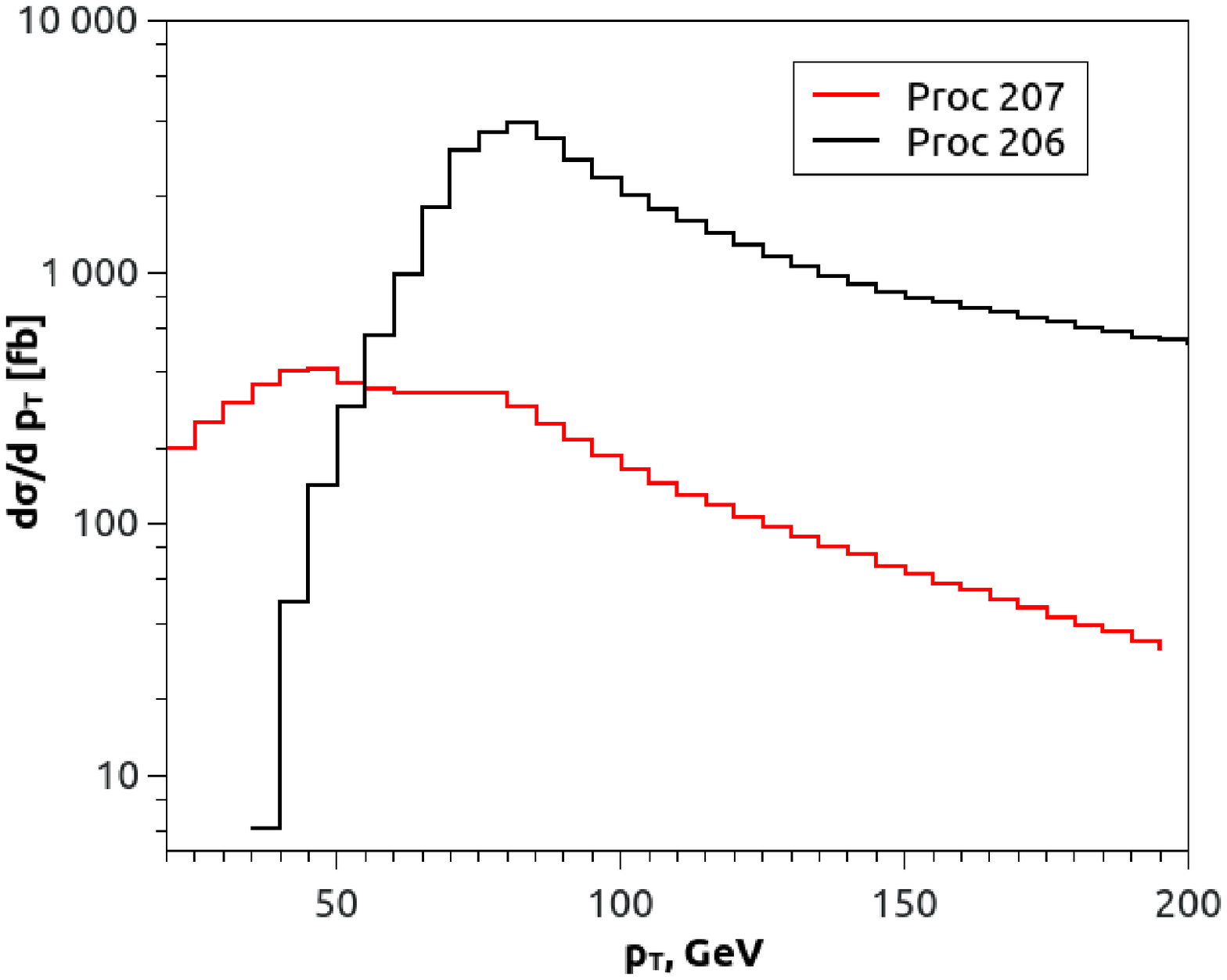}
 \includegraphics[width=0.45\textwidth]{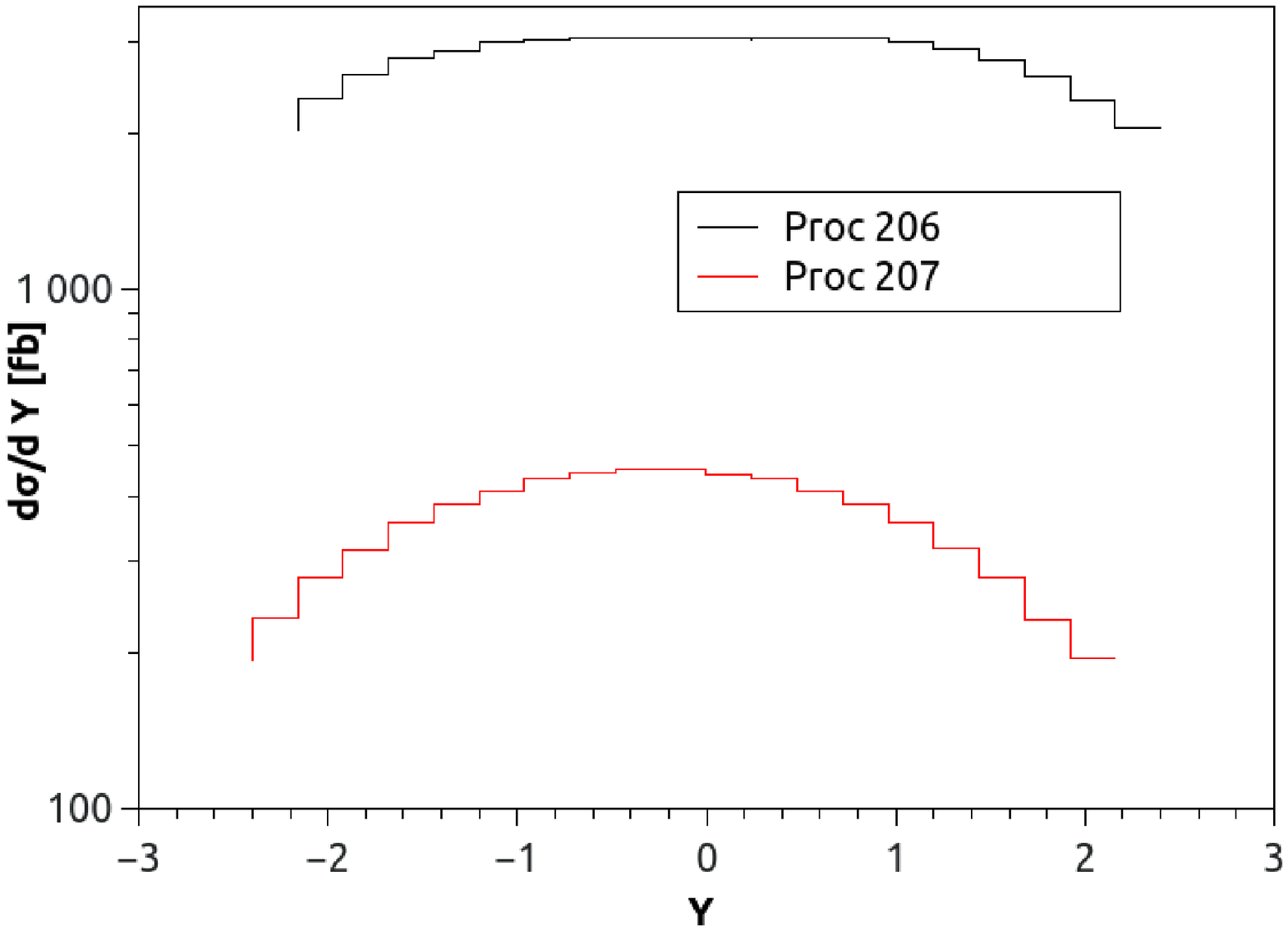}
 \caption{\label{fig:higgs_calc_pt_y} Differential production cross sections with respect to: (left) - A boson transverse momentum $p_T$; (right) A boson rapidity, $Y$.}
 \ec
\end{figure}

These processes represent the fermion interaction $f(p_1)+f(p_2)$ for the production of a Higgs boson A in association with a single jet $f(p_5)$, with the subsequent decay of A boson to either a pair of bottom quarks (206) or to a pair of tau’s (207). The calculated kinematical properties of A boson at 14 TeV for these two processes are presented in fig.\ref{fig:higgs_calc_pt_y}. 
From the comparision of 206 and 207 processes of the left part of fig.\ref{fig:higgs_calc_pt_y} we see the predominance of the first one and the different behavior at small transverse momenta. Right part of the fig.6 demonstrates also the predominance of 206 process contribution in the differential production cross section and its more slow drop in cross section when changing the angular distribution.

\par
Higgs self-coupling constant together with top-Yukawa coupling are most important parameters for studying of fundamental physics. Their accuracy is important for the studying of electroweak vacuum stability problem as well as for searches for new particles beyond SM. 

We will consider the following processes:\\
Process 640: neither  top quarks nor  Higgs boson decays;\\
Process 644: top quark decays leptonically and anti-top quark decays hadronically and Higgs boson decays into a pair of bottom quarks:\\
$t(\rightarrow \nu(p_3 ) + e^+(p_4) + b(p_5 )) + \overline{t}(\rightarrow \overline{q}(p_7 ) + q(p_8 ) + \overline{b}(p_6 )) + H(b(p_9 ) +  \overline{b}(p_{10} ))$\\
Such process is chosen due to the largest signal rate and in order to control the backgrounds using subjet method, \cite{DAWSON20191}. The calculation can be performed at LO only. 

\par
\begin{figure}[htbp]
\bec
 \includegraphics[width=0.45\textwidth]{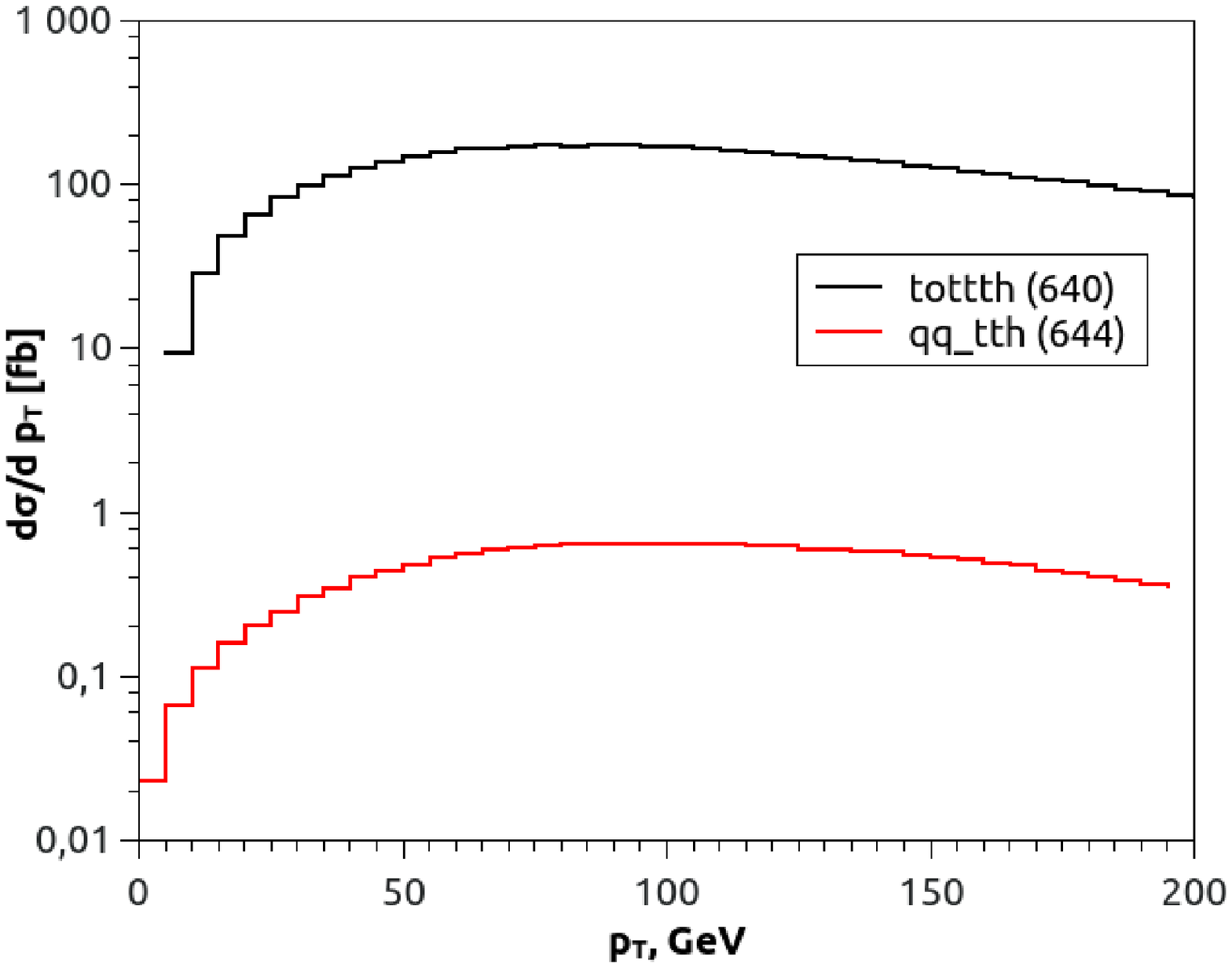}
 \includegraphics[width=0.45\textwidth]{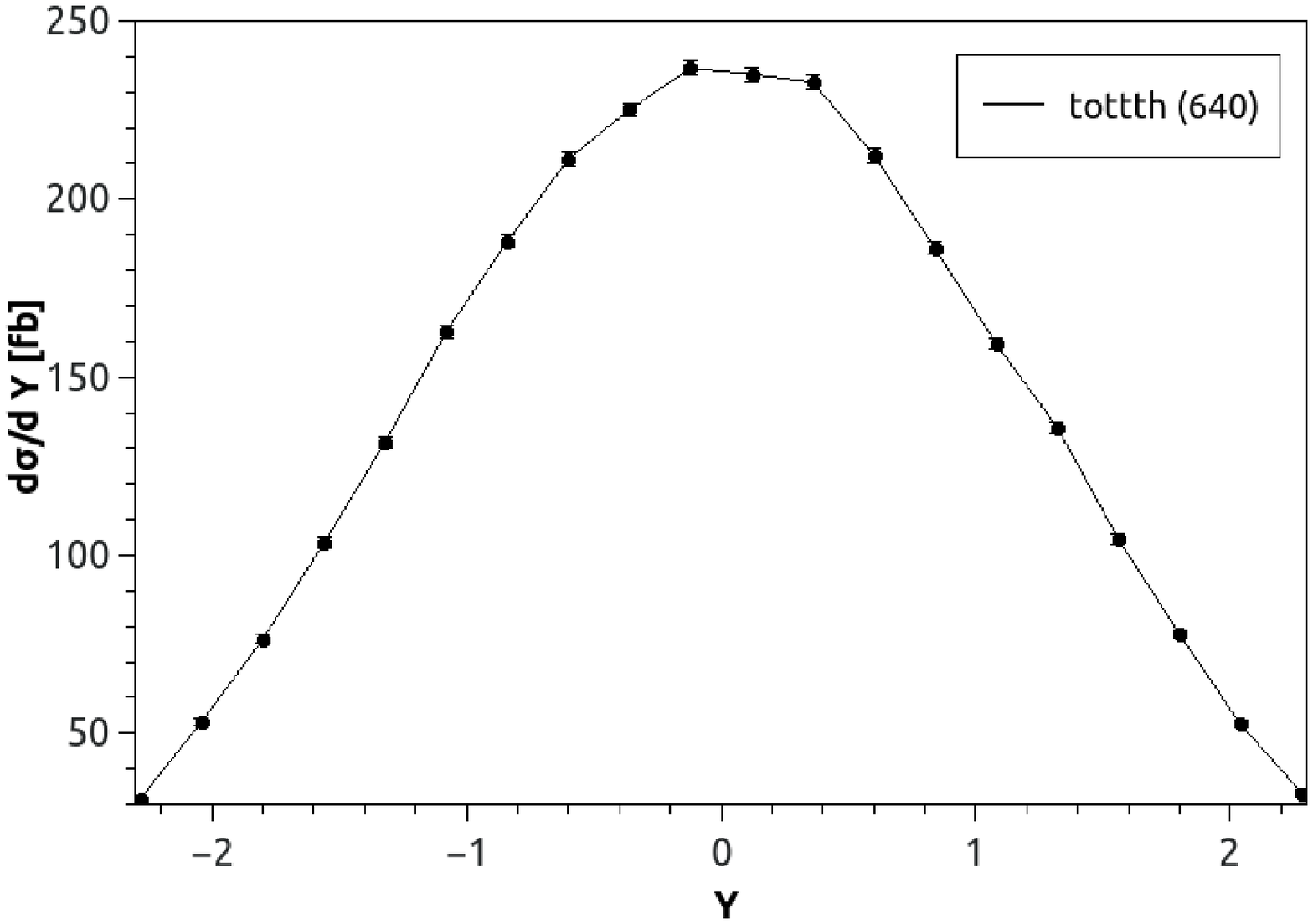}
 \caption{\label{fig:higgs_calc_pt_y_640} Differential production cross sections with respect to: (left) - Higgs boson transverse momentum $p_T$; (right) Higgs boson rapidity.}
 \ec
\end{figure}

In fig.\ref{fig:higgs_calc_pt_y_640} are presented QCD (for 640) and semi-hadronic (644) contributions into the production cross sections (left part) and rapidity distribution for Higgs boson (right part).
\par
From left part of fig.\ref{fig:higgs_calc_pt_y_640} we see the predominance of 640 process compared with 644 one due to the complex decay chain of the second process. As for the right-hand side of fig.\ref{fig:higgs_calc_pt_y_640}, it must be said that the maximum contribution into the production cross section of 640 process is from the processes perpendicular to the proton-proton collision axis.

\subsection{Dark matter production processes}
\par
As is known, approximately 85\% of the matter in the Universe is a form of matter called DM. The nature of DM is still unknown because its study is difficult due to the absence of any interaction other than gravitational one. So, one of the purposes of the LHC is to investigate the nature of DM in reactions with elementary particles of the following type:
\vspace{3mm}\\
842 '  $f(p1)+f(p2) \rightarrow S\rightarrow(X(p3)+X~(p4)) +f(p5)+f(p6)$ [Scalar Mediator] ' 'L'\\
843 '  $f(p1)+f(p2) \rightarrow PS\rightarrow(X(p3)+X~(p4)) +f(p5)+f(p6)$ [Pseudo Scalar Mediator] ' 'L'\\
844 '  $f(p1)+f(p2) \rightarrow GG\rightarrow(X(p3)+X~(p4)) +f(p5)+f(p6)$ [Gluonic DM operator] ' 'L'
\vspace{3mm}\\
Experimental searches for resonances -- DM mediators decaying into a pair of jets have been performed in proton-proton collisions at $\sqrt{s}$ = 13 TeV with integrated luminosity of 137 $fb^{-1}$ \cite{CMS-PAS-EXO-19-012}. DM mediators arise from an interaction between quarks and dark matter and have different nature. We will consider reactions with Scalar Mediator (S), Pseudo Scalar Mediator (PS) and Gluonic DM operator. Experimental restrictions on DM mediators below 2.8 TeV make it necessary to simulate them at higher energies at the LHC for different angular and momentum distributions. 
\par
\begin{figure}[htbp]
\bec
 \includegraphics[width=0.45\textwidth]{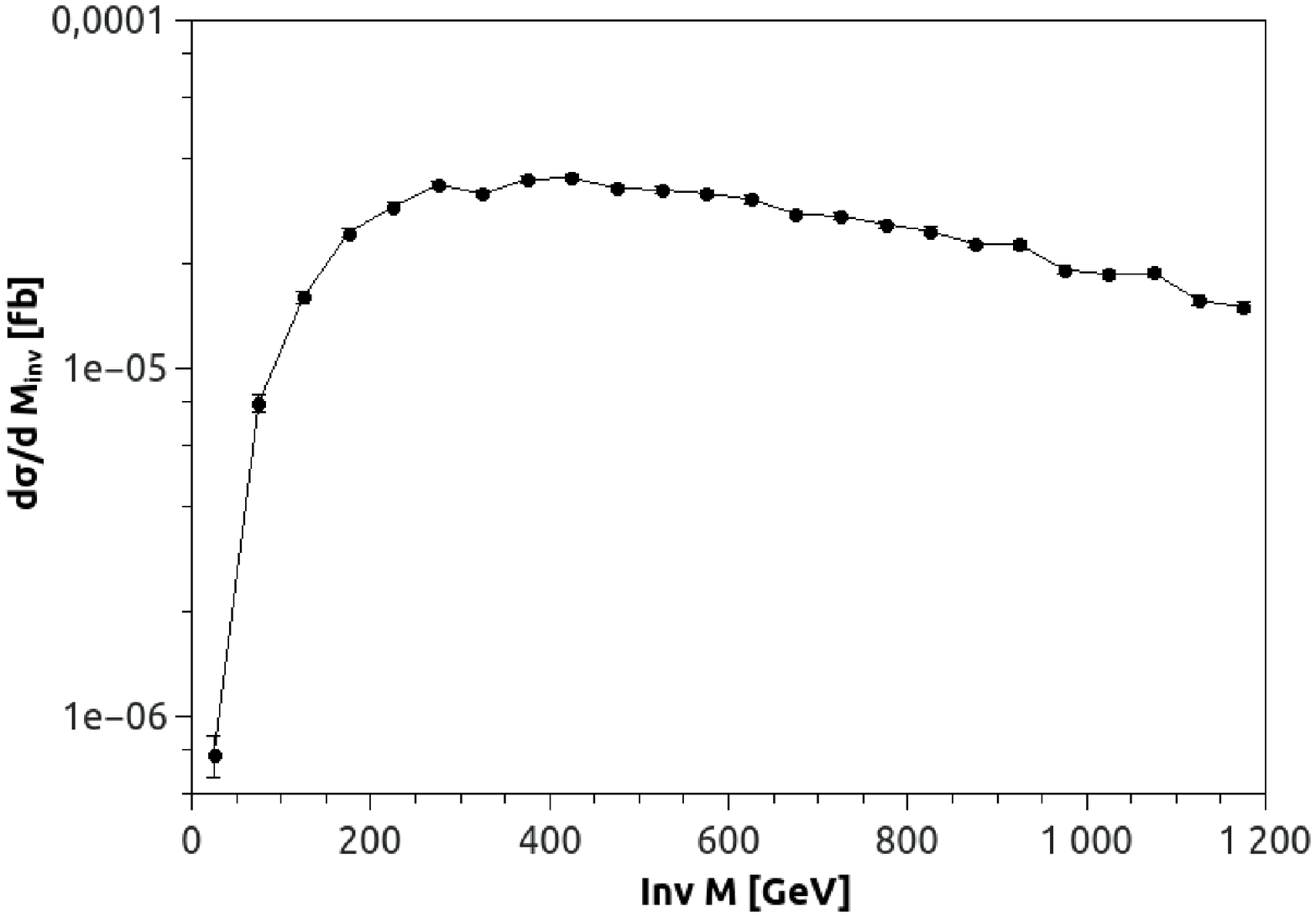}
 \includegraphics[width=0.43\textwidth]{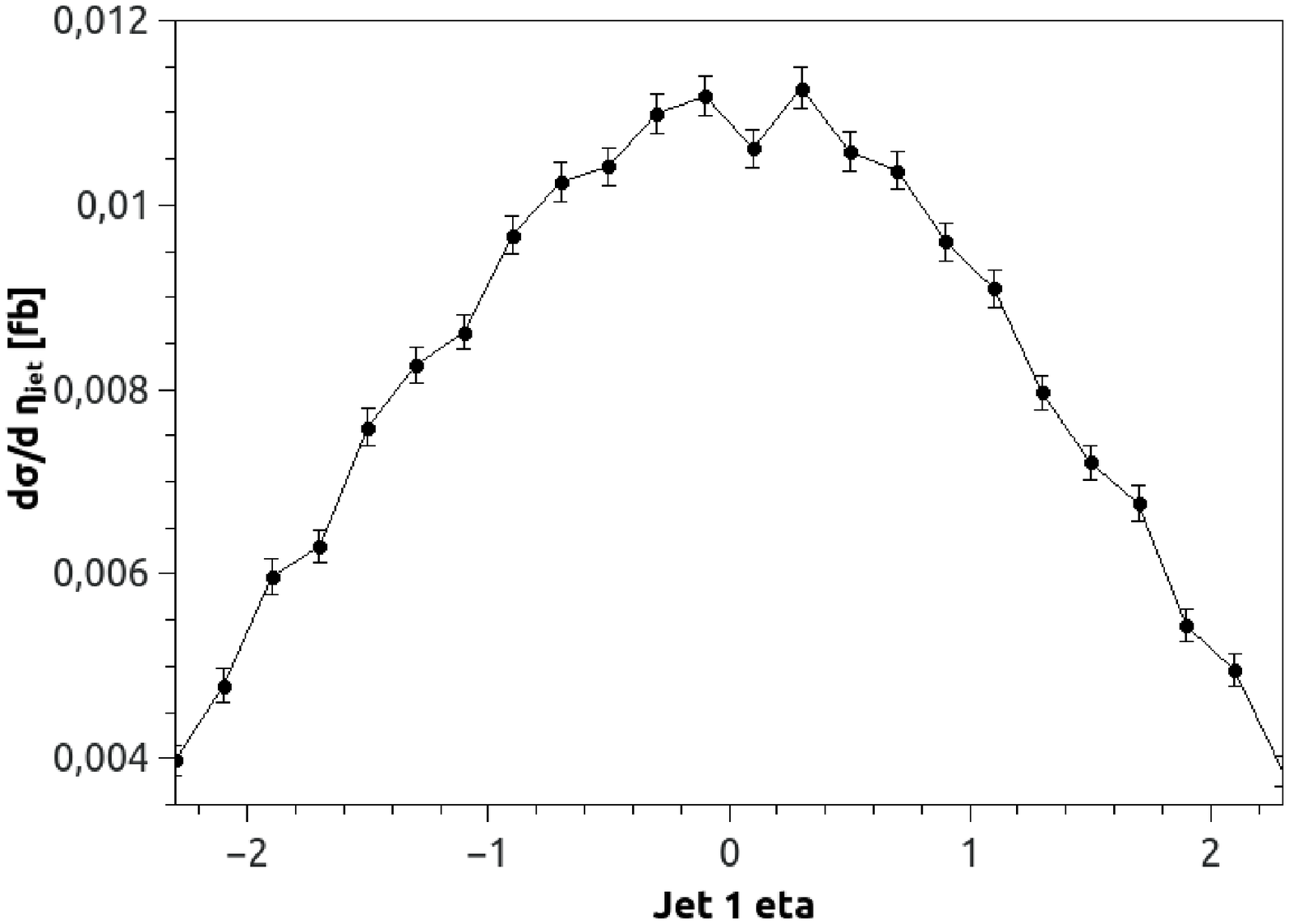}
 \includegraphics[width=0.45\textwidth]{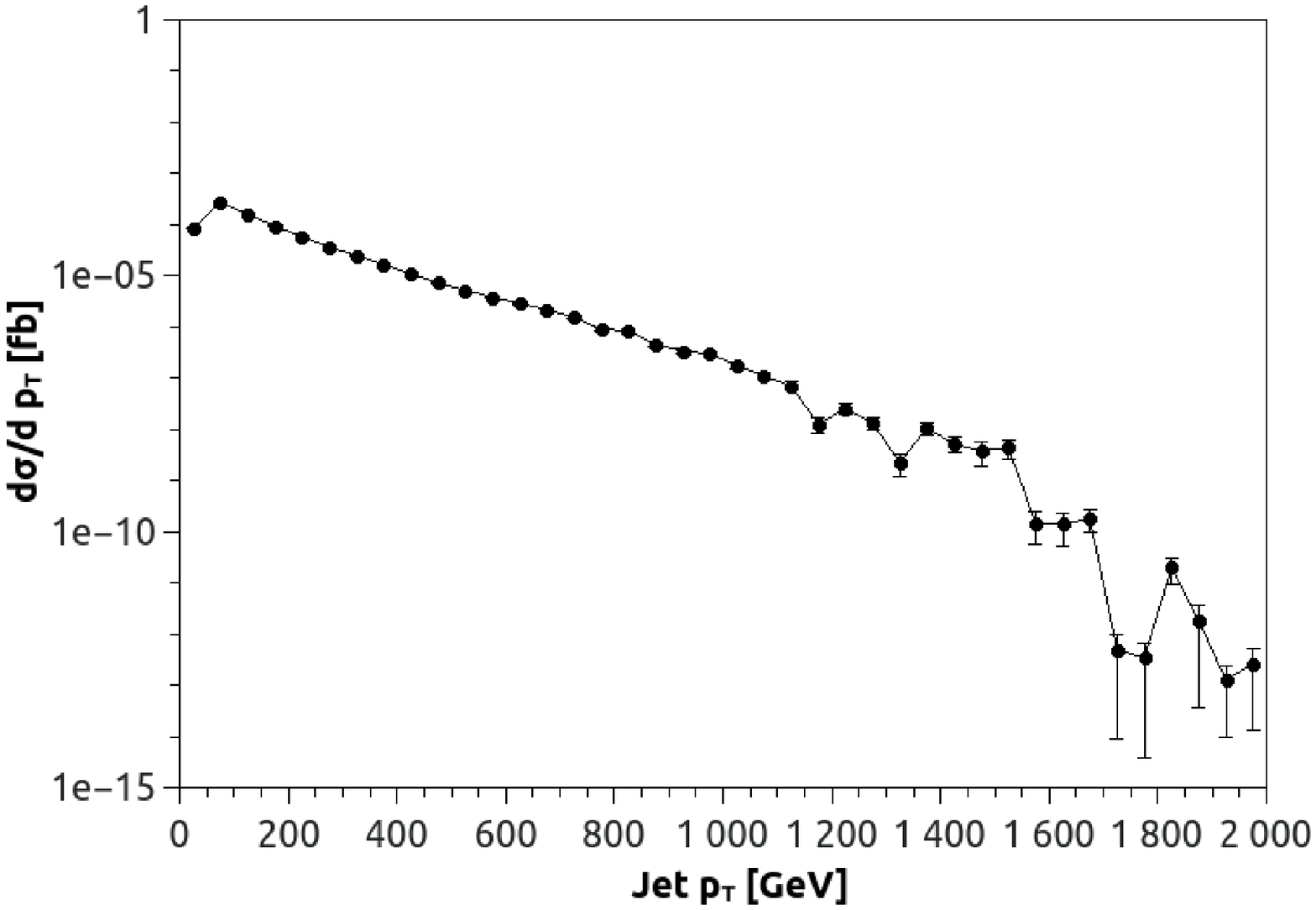}
 \caption{\label{fig:scalar_mediator} Scalar mediator invariant mass distribution (up and left), angular distribution (up and right) and momentum distribution (down) calculated for 14 TeV at the LHC.}
 \ec
\end{figure}

Using computer program MCFM v.9.0 we have calculated invariant mass distribution, angular and momentum distribution for process 842, which represent the production of DM plus two jets, fig.\ref{fig:scalar_mediator}.
\par
From fig.\ref{fig:scalar_mediator} we can see the maximal value of differential production cross section at about 200 GeV/c of invariant mass distribution. The angular distribution of the jet 1 is characterized by large, at about $85^{\circ}$ angle to the axis of the proton-proton interaction. Momentum distribution for differential production cross section shows the maximal value at about 100 GeV. An analysis of the complete kinematic information indicates about the mass region of S $\sim$ 200-400 GeV, the formation of which is accompanied by high-energy jets in the direction perpendicular to the axis of collision with $p_T > 100$ GeV. 

\par
\begin{figure}[htbp]
\bec
 \includegraphics[width=0.55\textwidth]{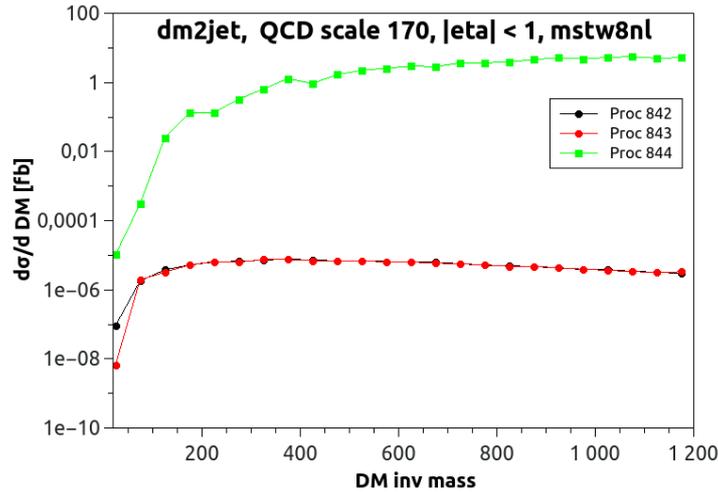}
 \caption{\label{fig:invarian_mass_842_844} Invariant mass distributions of 842-844 processes for 14 TeV in proton-proton collisions.}
\ec
\end{figure}

We also calculated invariant mass distributions for 842-844 processes in kinematic region $|\eta|<1$ and QCD scale = 170, presented in fig.\ref{fig:invarian_mass_842_844}.
\par
From comparision of these processes we see the predominance of the process with Gluonic DM operator and the absence of a clear resonance. The comparision of these data with Invariant mass distribution of Scalar Mediator in $|\eta|<2.4$, fig.\ref{fig:scalar_mediator}, shows that the difference between the maximal values of differential production cross section is about 1 order. This result indicates about the predominance of Scalar mediator process in the direction of perpendicular to the axis of collision.
\newpage
\section{Conclusions}

We have considered the following processes:

\begin{itemize}
\item Di-jet processes;
\item CP-odd A boson production process;
\item Higgs boson production process in association with pair of top quarks;
\item DM production processes.
\end{itemize}

These processes play the fundamental role for the searches of physics beyond the SM. With the help of computer program MCFM v.9.0 we have calculated differential production cross sections for invariant mass distribution and cinematic properties of the decay products at energy of 14 TeV. From the analysis of our results we made the following conclusions:

\begin{itemize}
\item the absence of a clear dependence on the Sudakov EW corrections both in the distribution in the invariant mass and in the momenta, but the need to take into account the angular distribution for the di-jet process;
\item the largest values of momentum and angular differential production cross sections for A boson are for the decay into b quarks compared to decay into tau leptons for $p_T > 80 GeV$. Thus the observation that A boson prefers to decay into heavier particles is confirmed;
\item complex decay chain of the Htt process, connected with leptonically decays of top quark and hadronically decays of anti-top quark is characterized by lower value of differential production cross section compared to the 640 process without any decays of Htt particles. Decay products of 640 process are oriented perpendicular to collision axis;
\item momentum distribution for differential production cross section of DM mediators shows the maximal value at about 100 GeV and complete kinematic information indicates about the mass region of Scalar Mediator of about 200-400 GeV. The formation of this DM mediator is accompanied by high-energy jets in the direction perpendicular to the axis of collision. Our calculations for three processes of formation of DM mediators shows the predominance of the process with Gluonic DM operator and the absence of a clear resonance.
\end{itemize}


\begin{thebibliography}{10}

\bibitem{CMS-PAS-EXO-19-012}
CMS Collaboration.``{A search for dijet resonances in proton-proton collisions at
  $\sqrt{s}=13~\mathrm{TeV}$ with a new background prediction method},'' Tech.
  Rep. CMS-PAS-EXO-19-012, CERN, Geneva, 2019.

\bibitem{Jarosik_2011}
N.~Jarosik, C.~L. Bennett, J.~Dunkley, B.~Gold, M.~R. Greason, M.~Halpern,
  R.~S. Hill, G.~Hinshaw, A.~Kogut, E.~Komatsu, D.~Larson, M.~Limon, S.~S.
  Meyer, M.~R. Nolta, N.~Odegard, L.~Page, K.~M. Smith, D.~N. Spergel, G.~S.
  Tucker, J.~L. Weiland, E.~Wollack, and E.~L. Wright,
  ``{SEVEN}-{YEARWILKINSON} {MICROWAVE} {ANISOTROPY} {PROBE}({WMAP})
  {OBSERVATIONS}: {SKY} {MAPS}, {SYSTEMATIC} {ERRORS}, {AND} {BASIC}
  {RESULTS},'' {\em The Astrophysical Journal Supplement Series}, vol.~192,
  p.~14, jan 2011.

\bibitem{PhysRevD.87.054030}
P.~J. Fox and C.~Williams, ``Next-to-leading order predictions for dark matter
  production at hadron colliders,'' {\em Phys. Rev. D}, vol.~87, p.~054030, Mar
  2013.

\bibitem{20121}
ATLAS Collaboration.``Observation of a new particle in the search for the standard model higgs
  boson with the ATLAS detector at the LHC,'' {\em Physics Letters B},
  vol.~716, no.~1, pp.~1 -- 29, 2012.

\bibitem{201230}
CMS Collaboration.``Observation of a new boson at a mass of 125 gev with the cms experiment at
  the LHC,'' {\em Physics Letters B}, vol.~716, no.~1, pp.~30 -- 61, 2012.

\bibitem{WESS197439}
J.~Wess and B.~Zumino, ``Supergauge transformations in four dimensions,'' {\em
  Nuclear Physics B}, vol.~70, no.~1, pp.~39 -- 50, 1974.

\bibitem{DIMOPOULOS1981150}
S.~Dimopoulos and H.~Georgi, ``Softly broken supersymmetry and su(5),'' {\em
  Nuclear Physics B}, vol.~193, no.~1, pp.~150 -- 162, 1981.

\bibitem{BRANCO20121}
G.~Branco, P.~Ferreira, L.~Lavoura, M.~Rebelo, M.~Sher, and J.~P. Silva,
  ``Theory and phenomenology of two-higgs-doublet models,'' {\em Physics
  Reports}, vol.~516, no.~1, pp.~1 -- 102, 2012.


\bibitem{ATLAS-CONF-2019-004}
ATLAS Collaboration.``{Measurement of Higgs boson production in association with a $t\overline t$
  pair in the diphoton decay channel using 139~fb$^{-1}$ of LHC data collected
  at $\sqrt{s} = 13$~TeV by the ATLAS experiment},'' Tech. Rep.
  ATLAS-CONF-2019-004, CERN, Geneva, Mar 2019.

\bibitem{ALEKHIN2012214}
S.~Alekhin, A.~Djouadi, and S.~Moch, ``The top quark and higgs boson masses and
  the stability of the electroweak vacuum,'' {\em Physics Letters B}, vol.~716,
  no.~1, pp.~214 -- 219, 2012.

\bibitem{campbell2019precision}
J.~Campbell and T.~Neumann, ``Precision phenomenology with MCFM,'' 2019.

\bibitem{1988127}
UA1 Collaboration.``Two-jet mass distributions at the CERN proton-antiproton collider,'' {\em
  Physics Letters B}, vol.~209, no.~1, pp.~127 -- 134, 1988.

\bibitem{bauce2017search}
M.~Bauce, ``Search for new physics in dijet final states in ATLAS and CMS,''
  2017.

\bibitem{CMS-PAS-EXO-19-004}
CMS Collaboration.``{Search for dijet resonances in events with three jets from proton-proton
  collisions at $\sqrt{s}=13~\mathrm{TeV}$},'' Tech. Rep. CMS-PAS-EXO-19-004,
  CERN, Geneva, 2019.

\bibitem{PhysRevD.94.093009}
J.~M. Campbell, D.~Wackeroth, and J.~Zhou, ``Study of weak corrections to
  drell-yan, top-quark pair, and dijet production at high energies with MCFM,''
  {\em Phys. Rev. D}, vol.~94, p.~093009, Nov 2016.

\bibitem{DAWSON20191}
S.~Dawson, C.~Englert, and T.~Plehn, ``Higgs physics: It ain’t over till it
  is over,'' {\em Physics Reports}, vol.~816, pp.~1 -- 85, 2019.

\end{thebibliography}
\end{document}